\documentclass[11pt,authoryear]{elsarticle}
\usepackage{mydef}
\usepackage{epstopdf}
\usepackage{tabularx}
\usepackage{enumitem}
\usepackage{subcaption}
\usepackage[show]{ed}
\usepackage{xcolor}
\addto\captionsenglish{\renewcommand{\appendixname}{Appendix }}

\def\hatr{{\hat r}}
\def\kapr{{\kappa_{\hat r}}}
\def\ther{{\theta_{\hat r}}}
\def\sigr{{\sigma_{\hat r}}}
\def\kapy{{\kappa_{y}}}
\def\they{{\theta_{y}}}
\def\sigy{{\sigma_{y}}}

\def\hatR{{\hat R}}
\def\hatB{{\hat B}}

\def\m1{\mathbbm{1}}
\def\calR{{\mathcal R}}
\def\calC{{\mathcal C}}
\def\calF{{\mathcal F}}
\def\QM{{\mathbb Q}}

\def\bfX{{\mathbf{X}_{t}}}

\def\bfX{{\mathbf{X}}}

\graphicspath{{./newFigs/}}
\begin{document}

\begin{frontmatter}

\title{Four-factor model of Quanto CDS \\ with jumps-at-default and stochastic recovery}
\markboth{A. Itkin, F. Soleymani}{...}

\author[nyu]{A.~Itkin\corref{cor1}}
\ead{aitkin@nyu.edu}
\cortext[cor1]{Corresponding author}

\author[Faz]{F.~Soleymani}
\ead{soleymani@iasbs.ac.ir}

\address[nyu]{Tandon School of Engineering, New York University, 1 MetroTech Center, 10 floor, Brooklyn NY 11201, USA}
\address[Faz]{Department of Mathematics, Institute for Advanced Studies in Basic Sciences (IASBS), Zanjan 45137-66731, Iran}

\begin{abstract}
In this paper we modify the model of \cite{isvIJTAF}, proposed for pricing Quanto Credit Default Swaps (CDS) and risky bonds, in several ways. First, it is known since the Lehman Brothers bankruptcy that the recovery rate could significantly vary right before or at default, therefore, in this paper we consider it to be stochastic. Second, to reduce complexity of the model, we treat the domestic interest rate as deterministic, because, as shown in \cite{isvIJTAF}, volatility of the domestic interest rate does not contribute much to the value of the Quanto CDS spread. Finally, to solve the corresponding systems of 4D partial differential equations we use a different flavor of the Radial Basis Function (RBF) method which is a combination of localized RBF and finite-difference methods, and is known in the literature as RBF--FD. Results of our numerical experiments presented in the paper demonstrate that the influence of volatility of the recovery rate is significant if the correlation between the recovery rate and the log-intensity of the default is non-zero. Also, the impact of the recovery mean-reversion rate on the Quanto CDS spread could be significant and comparable with the impact due to jump-at-default in the FX rate.
\end{abstract}

\begin{keyword}
Quanto CDS, reduced form model, jump-at-default, stochastic interest and recovery rates, RBF--FD method.

\JEL C51, C63, G13
\end{keyword}

\end{frontmatter}

\section{Introduction}
Quanto CDS is a flavor of a credit default swap. It has a special feature that in case of default
all payments including the swap premium and/or the cashflows are done not in a currency of the reference asset, but in another one.
As mentioned in \cite{Brigo}, a typical example would be a CDS that has its reference as a dollar-denominated bond for which the premium of the swap is payable in Euros. And in case of default the payment equals the recovery rate on the dollar bond payable in Euro.
This can also be seen as the CDS written on a dollar bond, while its premium is payable in Euro. These types of contracts are widely used to hedge holdings in bonds or bank loans that are denominated in a foreign currency (other than the investor home currency). In more detail, see, e.g., \cite{Brigo,isvIJTAF} and references therein.

\begin{figure}[!t]
\centering
\fbox{\includegraphics[width=0.8\textwidth]{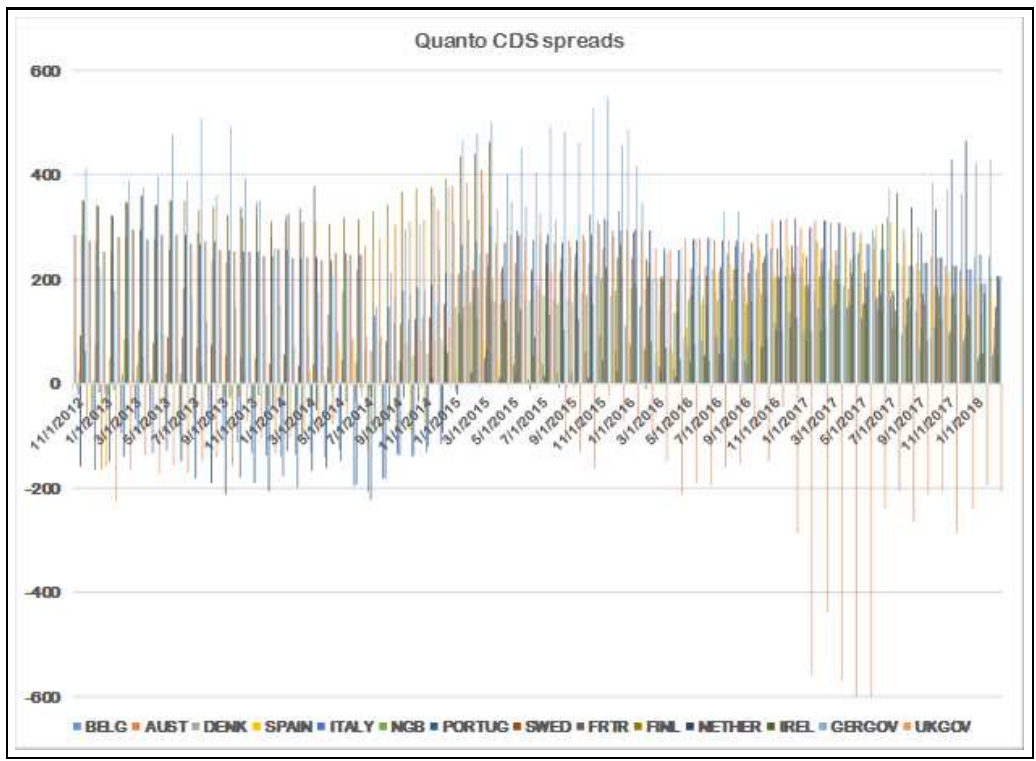}}
\caption{Quanto CDS spreads (in bps) computed based on historical time-series
data of some European 5Y sovereign CDS traded in both Euro and USD.}
\label{histSpread}
\end{figure}

To illustrate, in Fig.~\ref{histSpread} Quanto CDS spreads computed by using historical time series\footnote{The CDS data are from Markit.} of some European 5Y sovereign CDS traded in both Euro and USD are presented for the period from 2012 to January 2018. It can be seen that nowadays these spreads still could reach 400 bps and more for some countries, and thus, building a comprehensive model capable to predict quanto effects remains a problem of current interest.

In \cite{ACS2017} Quanto CDS spreads, defined as the difference between the USD and EUR denominated CDS spreads, are presented for six countries from Eurozone: Belgium, France, Germany, Ireland, Italy, Portugal, at maturities 3, 5, 7, 10, and 15 years relative to the 1 year quanto spread. This data demonstrate the spreads reaching 30 bps at 15 years (France, Ireland). In \cite{Simon2015} the 5 years Quanto CDS spreads are reported for Germany, Italy and France over the period from 2004 to 2013, which, e.g., for Italy could reach 500 bps in 2012. The results presented in \cite{Brigo} indicate a significant basis across domestic and foreign CDS quotes. For instance, for Italy a USD CDS spread quote of 440 bps can translate into a EUR quote of 350 bps in the middle of the Euro-debt crisis in the first week of May 2012. More recent basis spreads (from June 2013) between the EUR quotes and the USD quotes are in the range of 40 bps.

On a modeling site \cite{Brigo} proposed a model for pricing Quanto CDS based on the reduced form model for credit risk. They represent the default time as a Cox process with explicit diffusion dynamics for the default intensity/hazard rate and an exponential jump to default, similar to the approach of \cite{ES2006, Mohammadi2006}. Also they introduce an explicit jump-at-default in the FX dynamics. This provides a more efficient way to model credit/FX dependency as the results of simulation are able to explain the observed basis spreads during the Euro-debt crisis. While accounting only for the instantaneous correlation between the driving Brownian motions of the default intensity and the FX rate alone is not sufficient for doing so.

In \cite{isvIJTAF} the framework of \cite{Brigo} was extended by introducing stochastic interest rates and jump-at-default in both FX and foreign (defaulted) interest rates. The authors  investigated a relative contribution of both jumps into the magnitude of the Quanto CDS spread. The results presented in the paper qualitatively explain the discrepancies observed in the marked values of CDS spreads traded in domestic and foreign
economies and, accordingly, denominated in the domestic (USD) and foreign (Euro, Ruble, Brazilian Real, etc.) currencies. The quanto effect (the difference between the prices of the same CDS contract traded in different economies, but represented in the same currency) can, to a great extent, be explained by the devaluation of the foreign currency. This  would yield a much lower protection payout if converted to the US dollars. These results are similar to those obtained in \cite{Brigo}. However, in \cite{Brigo} only constant foreign and domestic interest rates are considered, while in \cite{isvIJTAF} they are stochastic even in the no-jumps framework.

In contrast to \cite{Brigo}, in \cite{isvIJTAF} the impact of the jump-at-default in the foreign interest rate, which could occur simultaneously with the jump in the FX rate, was also analyzed. The authors found that this jump is a significant component of the process and is able to explain about 20 bps of the basis spread value. They also show that the jumps in the FX rate $z$ and the foreign interest rate $\hatr$ have opposite effects. In other words,  devaluation of the foreign currency will decrease the value of the foreign CDS, while the increase of $\hatr$ will increase the foreign CDS value. Influence of other important model parameters such as correlations between the log-hazard rate $y$ and the factors that incorporate jumps, e.g., $\rho_{yz}$ and $\rho_{y\hatr}$, and volatilities of the log-hazard process $\sigma_y$ and the FX rate $\sigma_z$ was also investigated in \cite{isvIJTAF}, with the conclusion made that they have to be properly calibrated. The other correlations just slightly contribute to the basis spread value. Large values of the volatilities can in some cases explain up to 15 bps of the basis spread value.

Despite, the model of \cite{isvIJTAF} is sufficiently complex and capable to qualitatively explain the behavior of the Quanto CDS spreads observed in the market, it could be further improved in several ways. First, it is known since the Lehman Brothers bankruptcy that the recovery rates could significantly vary right before or at default. For instance, in \cite{Brigo2015} this is modeled by considering a time-dependent (piece-wise constant) recovery rate. However, such an approach does not take into account possible correlations of the recovery rate with the other stochastic drivers of the model. Meanwhile, in the literature the existence of such correlations is reported, especially for the correlation between the default (hazard) and recovery rates, \cite{Altman2002}. For instance, in \cite{Witzany2013} a two-factor model of \cite{Rosch2009} is used to capture the recovery rate variation and its correlation with the rate of default. The author suggest two approaches to estimate the model parameters: based on time-series of the aggregate default and recovery rates, and a cross-sectional approach based on the exposure level data. The results (based on the Moody's DRD database) confirm not only significant variability of the recovery rate, but also a significant correlation over 50\% between the rate of default and the recovery rates in the context of the model.

Therefore, in this paper we let the recovery rate to be stochastic. At the same time, to reduce complexity of the model, we relax the assumption of \cite{isvIJTAF} about stochasticity of the domestic interest rate. This is  because, as shown in \cite{isvIJTAF}, the volatility of the domestic interest rate does not contribute much to the value of the Quanto CDS spread. Thus, still our model has four stochastic drivers, as in \cite{isvIJTAF}.

Another modification as compared with \cite{isvIJTAF} is related to the numerical method used to compute the CDS prices. Since the pricing problems in \cite{isvIJTAF} are formulated via backward partial differential equations (PDE), computation of the CDS price for every maturity $T$ requires solving as many independent backward problems as the number of time steps from $0$ to $T$ is chosen. This could be improved in two ways. The first option would be if instead of the backward PDE we would work with the forward one for the corresponding density function. In that case all solutions  $U_t(t_i), \ i \in [0,m], \ t_0 = 0, t_m = T$ for all times $t_i$ can be computed in one sweep, i.e., by a marching method. This situation is similar to that in option pricing, where the forward equation is useful for calibration as it allows computation of the option smile in one sweep\footnote{In other words, given the spot price the option prices for multiple strikes and maturities could be computed by solving just one forward equation.}. In contrast, solving the backward equation is useful for pricing, as given the strike it allows computation of the option prices for various initial spot values in one sweep, see, e.g., \cite{Itkin2014b}.
However, we do not use this approach in this paper, and our results obtained by using the forward approach will be reported elsewhere.

The other way to significantly accelerate calculations assumes that still, the backward equations need to be solved, however this can be done by using parallelization of calculations. Indeed, as we demonstrate it below in the paper, solutions  $U_t(t_i), \ i \in [0,m], \ t_0 = 0, t_m = T$ for all times $t_i$ can be found in parallel. This is our approach in hands for this paper.

Finally, for solving our systems of PDEs we use a different flavor of the Radial Basis Function (RBF) method\footnote{In \cite{isvIJTAF} a flavor of the RBF-PUM method is used.} which is a combination of localized RBF and finite-difference methods (see \cite{SM_RBF2017,FLYER201621, Bayona2017} and references therein), and is known in the literature as the RBF--FD method. More specifically, a flavor of the RBF--FD method used in this paper is described in
\cite{Soleymani2018,FazItkin2019}. Both the RBF-PUM and RBF--FD methods belong to a localized version of the classical RBF, and demonstrate similar features, such as high accuracy, sparsity of the differentiation matrices, mesh-free nature and multi-dimensional extendability.
The comparison of these methods presented in \cite{SM_RBF2017}  illustrates capability of both methods to solve the problems to a sufficient accuracy with reasonable time, while both methods exhibit similar orders of convergence, However, from the potential parallelization viewpoint the RBF--FD is, perhaps, more suitable.

The rest of the paper is organized as follows. In Section~\ref{model} we describe our model, and derive the main PDEs for the risky bond price under this model. We first introduce a no-jumps framework, and then  extend this framework by adding jumps-at-default into the dynamics of the FX and foreign  (defaulted) interest rates. In Section~\ref{zcbPrice} we describe a backward PDE approach for pricing zero-coupon bonds. The connection of this price with the price of the Quanto CDS is established in Section~\ref{bond2cds}. In Section~\ref{numMethod} the RBF--FD method is described in detail. In Section~\ref{experiments} we present numerical results of our experiments obtained by using this model and discuss the influence of various model parameters on the basis quanto spread. Section~\ref{sec:Conclusion} concludes.

\section{Model} \label{model}
Below we describe our model following the same notation and definitions as in \cite{isvIJTAF}.

Let us denote the most liquidly traded currency among  all contractual currencies as the {\it domestic currency} or the {\it liquid currency}. In this paper this is the US dollars (USD). We also denote the other contractual currency as {\it contractual} or {\it foreign currency}. In this paper it can be both, e.g., EUR, and USD. Payments of the premium and protection legs of the contract are settled in this currency. We assume that CDS market quotes are available in both currencies (domestic and foreign), and denote these prices as $\mathrm{CDS}_d$ and $\mathrm{CDS}_f$ respectively.

The price $\mathrm{CDS}_f$ can be alternatively expressed in the domestic currency if the exchange rate $Z_t$ for two currencies is given by the market. This implies that the price of the CDS contract denominated in the domestic currency could be also expressed in the foreign currency as $Z_t \mathrm{CDS}_d$. If these two prices are different, one can introduce a spread $\mathrm{CDS}_f - Z_t \mathrm{CDS}_d$. It is known that this spread implied from the market could reach hundreds of bps, \cite{Brigo,ACS2017,Simon2015}. Thus,
these spreads could be detected if the market quotes on the CDS contracts in both currencies and the corresponding exchange rates are available.

Further on, as the risk neutral probability measure $\mathbb{Q}$ we choose that corresponding to the domestic (liquid) currency money market. Also, by $\mathbb{E}_t[\,\cdot\,]$ we denote the expectation conditioned on the information received by time $t$, i.e. $\mathbb{E}[\,\cdot\, | \mathcal{F}_t]$. We also denote a zero-coupon bond price associated with the domestic currency (USD) as $B_t$, and that  associated with the foreign currency (EUR) as $\hatB_t$, where $t\geq 0$ is the calendar time. We assume that the dynamics of these two money market accounts is given by
\begin{alignat}{2} \label{mmDyn}
dB_t & = r(t) B_t dt, \quad &&B_0 = 1,\\
d\hatB_t & = \hatr_t \hatB_t dt, \quad &&\hatB_0 = 1. \nonumber
\end{alignat}
Here $r(t)$ is the deterministic domestic interest rate, and $\hatr_t$ is the stochastic foreign interest rate. As compared with the model setting in \cite{isvIJTAF}, in this paper $r(t)$ is assumed to be deterministic while in \cite{isvIJTAF} it is stochastic. However, as it has been already explained, this is done because, based on the results of \cite{isvIJTAF}, the volatility of the domestic interest rate does not contribute much to the value of the Quanto CDS spread.

Similar to \cite{isvIJTAF}, first we consider a setting where all underlying stochastic processes do not experience  jump-at-default except the default process itself. Then in Section~\ref{modelJumps} this will be generalized by taking into account jumps-at-default in other processes.

\subsection{No jumps-at-default}

We assume that $\hatr_t$ follows the Cox-Ingersoll-Ross (CIR) process, \cite{cir:85}
\begin{align} \label{dynR}
  d\hatr_t &= \kapr(\ther - \hatr_t) dt + \sigr \sqrt{\hatr_t}dW_t^{(2)}, \quad \hatr_0=\hatr,
\end{align}
where $\kapr, \ther$ are the mean-reversion rate and level, $\sigr$ is the volatility, $W_t^{(2)}$ is the Brownian motion, and $\hatr$ is the initial level of the foreign interest rate. Without loss of generality, in this paper we assume $\kapr, \ther, \sigr$ to be constant, while this can be easily relaxed to have them time-dependent.

The exchange rate $Z_t$ denotes the amount of domestic currency one has to pay to buy one unit of foreign currency (so 1 Euro could be exchanged for $Z_t$ US dollars).
It is assumed to be stochastic and follow a log-normal dynamics
\begin{equation} \label{Z}
dZ_t = \mu_z Z_t dt + \sigma_z Z_t dW_t^{(3)}, \quad Z_0=z,
\end{equation}
\noindent where $\mu_z, \sigma_z$ are the corresponding drift and volatility, and $W_t^{(3)}$ is another Brownian motion.

As the underlying security of a CDS contract is a risky bond, we need a model of the credit risk  implied by the bond.
Here we rely on a reduced form model approach, see e.g., \cite{jarrow/turnbull:95, DuffieSingleton99, Bielecki2004, jarrow2003robust} and references therein. We define the hazard rate $\lambda_t$ to be the exponential  Ornstein-Uhlenbeck process
\begin{align} \label{lambda}
\lambda_t &= e^{Y_t}, \quad t \ge 0, \\
dY_t &= \kapy(\they-Y_t)dt + \sigy dW_t^{(4)}, \quad Y_0=y, \nonumber
\end{align}
\noindent where $\kapy, \they, \sigy$ are the corresponding mean-reversion rate, the mean-reversion level and the volatility, $W_t^{(4)}$ is the Brownian motion, and $y$ is the initial level of $Y$. Both $Z_t$ and $\lambda_t$  are defined and calibrated in the domestic measure.

In contrast to \cite{Brigo,isvIJTAF} in this paper we let the recovery rate to be stochastic. It is popular in the financial literature and also among rating agencies such as Moody’s to model the recovery rate using the Beta distribution. The reasons behind that and an extended survey of the existing literature on the subject can be found, e.g., in \cite{Morozovskiy2007}. As it is well-known, Beta distributions are a family of continuous time distributions defined on the interval $[0,1]$. Since values of recoveries also fall to the same interval, the domain of the Beta distribution can be viewed as a recovery value. The pdf of the Beta distribution is, \cite{BetaDistrib2001}
\begin{equation} \label{beta}
f(R, \alpha, \beta) = \frac{\Gamma(\alpha+\beta)}{\Gamma(\alpha)\Gamma(\beta)} \calR^{\alpha-1} (1-\calR)^{\beta-1}, \quad \calR \in [0,1],
\end{equation}
\noindent where $\calR$ is the recovery rate, $\Gamma(x)$ is the gamma function, and $\alpha > 0, \beta > 0$ are shape parameters.

However, to be compatible with our general setting where the dynamics of the stochastic drivers $Z_t, \hatr_t, Y_t$ is described by stochastic differential equations (SDE), and introduce correlations between them, it would be convenient also to introduce some SDE which the Beta distribution solves. For instance, following \cite{Flynn2004} we can write
\begin{equation} \label{betaSDE}
d\calR = \kappa_R (\theta_R - \calR) dt + \sigma_R \sqrt{\calR(1-\calR)} d W_t^{(1)}, \quad \calR_0 = R.
\end{equation}
Here $\kappa_R, \theta_R$ are the mean-reversion rate and level, $\sigma_R$ is the volatility of the recovery rate, $W_t^{(1)}$ is the Brownian motion, and $R$ is the initial level of the recovery rate. It is known, that the stationary solution for this SDE is the Beta probability density function, given in \eqref{beta} with
\begin{equation} \label{betaParams}
\alpha = \frac{\kappa_R \theta_R}{\sigma_R^2}, \quad \beta = \frac{\kappa_R (1-\theta_R)}{\sigma_R^2}.
\end{equation}

We assume all Brownian motions $W_t^{(i)}, \ i \in [1,4]$ to be correlated with the constant instantaneous correlation $\rho_{ij}$ between each pair $(i,j)$: $\langle d W_t^{(i)}, d W_t^{(j)} \rangle = \rho_{ij} dt$. Thus, the whole correlation matrix reads
\begin{equation}
\mathfrak{R} =
 \begin{bmatrix}
    1 & \rho_{R \hatr} & \rho_{Rz} & \rho_{Ry} \\
    \rho_{\hatr R} & 1 & \rho_{\hatr z} & \rho_{\hatr y} \\
    \rho_{zR} & \rho_{z \hatr} & 1 & \rho_{z y} \\
    \rho_{yR} & \rho_{y \hatr} & \rho_{yz} & 1 \\
  \end{bmatrix},
\qquad
|\rho_{ij}| \le 1, \quad i,j \in [R,\hatr, z, y].
\end{equation}

The default process $(D_t, \ t \ge 0)$ is defined as
\begin{equation} \label{defProc}
D_t = {\bf 1}_{\tau \le t},
\end{equation}
\noindent where $\tau$ is the default time of the reference entity. In order to exclude trivial cases, we assume that $\QM(\tau > 0) = 1$, and  $\QM(\tau \le T) > 0$.

\subsection{A jump-at-default framework} \label{modelJumps}

Following \cite{isvIJTAF}, we now extend our model by assuming that the exchange rate and the foreign interest rate can jump at the default time. The jump in the exchange rate is due to denomination of the foreign currency which was observed during the European sovereign debt crisis of 2009-2010.

It is shown in \cite{Brigo} that allowing for jump-at-default in the FX rate provides a way of modeling the credit/FX dependency which is capable to explain the observed basis spreads during the Euro-debt crisis. In contrast, another approach which takes into account only instantaneous correlations imposed among the driving Brownian motions of the default intensity and FX rates is not able to do so. Then, in \cite{isvIJTAF} it was argued that an existence of jump-at-default in the foreign interest rate could also be justified by historical time-series, especially in case when sovereign obligations are in question. Therefore, in \cite{isvIJTAF} both jump-at-default in the FX and foreign interest rates were considered, and their relative contribution to the value of the Quanto CDS spread was reported.

In this paper we reuse this approach. Namely, to add jumps to the dynamics of the FX rate in \eqref{Z}, we follow \cite{Brigo, BieleckiPDE2005} who assume that at the time of default the FX rate experiences a single jump which is proportional to the current rate level, i.e.
\begin{equation} \label{jumpZ}
d Z_t = \gamma_z Z_{t^-} d M_t,
\end{equation}
\noindent where $\gamma_z \in [-1,\infty)$ \footnote{This is to prevent $Z_t$ to be negative, \cite{BieleckiPDE2005}.} is a devaluation/revaluation parameter.

The hazard process $\Gamma_t$ of a random time $\tau$ with respect to a reference filtration is defined through the equality $e^{-\Gamma_t} = 1 - \QM\{\tau \le t|\calF_t\}$. It is well known that if the hazard process $\Gamma_t$ of $\tau$ is absolutely continuous, so
\begin{equation} \label{hazard}
\Gamma_t = \int_0^t (1-D_s) \lambda_s ds,
\end{equation}
\noindent and increasing, then the process $M_t = D_t - \Gamma_t$ is a martingale
(which is called as the compensated martingale of the default process $D_t$) under the full filtration $\calF_t \vee {\mathcal H}_t$ with ${\mathcal H}_t$ being the filtration generated by the default process. So, $M_t$ is a martingale under $\QM$, \cite{BieleckiPDE2005}.

It can be shown that under the risk-neutral measure associated with the domestic currency, the drift $\mu_z$ is, (\cite{Brigo})
\begin{equation} \label{na-drift}
\mu_z = r(t) - \hatr_t.
\end{equation}

Therefore, with the allowance for \eqref{Z}, \eqref{jumpZ} we obtain
\begin{equation} \label{dzJump}
dZ_t =  [r(t) - \hatr_t] Z_t dt + \sigma_z Z_t dW_t^{(3)} + \gamma_z Z_t d M_t.
\end{equation}
Thus, $Z_t$ is a martingale under the $\mathbb{Q}$-measure with respect to $\calF_t \vee {\mathcal H}_t$ as it should be, since it is a tradable asset.

As the default of the reference entity is expected to negatively impact the value of the local currency, further on
we consider mostly the negative values of $\gamma_z$. For instance, we expect the value of EUR expressed in USD to fall down in case if some European country defaults.

Similarly, we add jump-at-default to the stochastic process for the foreign interest rate $\hatr_t$ as
\[ d \hatr_t = \gamma_\hatr \hatr_{t^-} d D_t, \]
\noindent so \eqref{dynR} transforms to
\begin{equation} \label{rJump}
d\hatr_t = \kapr(\ther-\hatr_t )dt + \sigr \sqrt{\hatr_t}dW_t^{(2)} + \gamma_{\hatr} \hatr_t d D_t.
\end{equation}
Here $\gamma_{\hatr} \in [-1,\infty)$ is the parameter that determines the post-default cost of borrowing. In this paper we consider just the positive values of $\gamma_{\hatr}$, because most likely the interest rate grows after the default has occurred\footnote{Here we consider sovereign CDS, so the default is associated with the national economy. Therefore, further evolution of the foreign interest rate is, perhaps, subject to government actions. However, the scenario in which the interest rate will increase after the default seems to be preferable.}. Note that $\hatr_t$ is not tradable, and so is not a martingale under the $\mathbb{Q}$-measure.

\section{PDE approach for pricing zero-coupon bonds} \label{zcbPrice}

Here we consider a general building block necessary to price quanto contingent claims, i.e., claims where the contractual currency differs from the pricing currency,  such as Quanto CDS. For doing so we determine a price of a defaultable zero-coupon bond
settled in the foreign currency. Under the foreign money market martingale measure $\hat \QM$,
this bond price reads
\begin{equation}
 \hat U_t(T) = \hat{\mathbb{E}}_t\left[ \frac{\hatB_t}{\hatB_T} \hat \Phi(T) \right],
\end{equation}
\noindent where $\hatB_t/\hatB_T = \hat B(t,T)$ is the stochastic discount factor from time $T$ to time $t$ in the foreign economy, and $\Phi(T)$ is the payoff function.
However, to compute the quanto spread we need this price under the domestic money market measure $\mathbb{Q}$. It can be obtained by converting the payoff to the domestic currency and discounting by the domestic money market account that yields
\begin{equation}
  U_t(T) = \mathbb{E}_t\left[ B(t,T) Z_t \hat \Phi(T) \right].
\end{equation}
Here it is assumed that the notional amount of the contract is equal to one unit of the foreign currency, hence the payoff function is
\begin{equation} \label{payoff1}
  \hat \Phi(T) = \m1_{\tau>T}.
\end{equation}
Let us assume that in case of the bond default, the recovery rate $\calR$ is paid at the default time. Then the price of a defaultable zero-coupon bond, which pays out one unit of the foreign currency in the domestic economy reads
\begin{align} \label{payoff}
U_t(T) &= \mathbb{E}_t\left[ B(t,T) Z_T \m1_{\tau>T}
+ \calR_\tau B(t,\tau) Z_\tau \m1_{\tau \le T} \right] \\
&= \mathbb{E}_t\left[ B(t,T) Z_T \m1_{\tau>T} \right]
+  \int_t^T \mathbb{E}_t \left[ \calR_\nu B(t,\nu) Z_\nu \m1_{\tau \in (\nu-d\nu,nu]} \right]  = w_t(T) + \int_t^T g_t(\nu)d \nu, \nonumber \\
w_t(T) &:= \mathbb{E}_t \left[ Z_{T} B(t,T) \m1_{\tau > T} \right], \qquad
g_t(\nu) := \mathbb{E}_t \left[ \calR_\nu B(t,\nu) Z_\nu \frac{\m1_{\tau \in (\nu-d\nu,nu]}}{d\nu} \right]. \nonumber
\end{align}

As the entire dynamics of the underlying processes is Markovian, \cite{BieleckiPDE2005}, finding the price of a defaultable zero-coupon bond can be done by using a PDE approach, i.e., this price just solves this PDE. This is more efficient from the computationally point of view as compared, e.g., with the Monte Carlo method, despite the resulting PDE becomes four-dimensional.

Since in this paper we prefer to use a backward approach, this PDE can be obtained by first conditioning on $\calR_t = R, \hatr_t = \hatr, Z_t = z, Y_t = y, D_t = d$ \footnote{Here $d$ can take two values: 0, which means that the default is not occurred yet, and 1 which means that the default has already occurred.}, and then using the approach of \cite{BieleckiPDE2005}  (a detailed derivation can be found in Appendix of \cite{isvIJTAF}). Under the risk-neutral measure $\QM$ the price $U_t(T)$ then reads
\begin{equation} \label{bondPrice}
U_t(T, R, \hatr, y, z) = \m1_{\tau > t} f(t, T, R,\hatr, y, z, 0) +
\m1_{\tau \le t} f(t, T, R,\hatr, y, z, 1).
\end{equation}
The function $f(t, T, R,\hatr, y, z, 1) \equiv u(t, T, X), \ \bfX := \{R,\hatr, y, z\}$ solves the PDE
\begin{equation} \label{PDE1}
\fp{u(t,T,\bfX)}{t} + {\cal L} u(t,T,\bfX) - r u(t,T,\bfX) = 0,
\end{equation}
\noindent and $\cal L$ is the diffusion operator which reads
\begin{align} \label{Ldiff}
\cal L &= \frac{1}{2}\sigma_{R}^2 R(1-R)\sop{}{R} + \frac{1}{2} \sigma_{\hatr}^2 \hatr \sop{}{\hatr} + \frac{1}{2}\sigma_z^2 z^2 \sop{}{z} + \frac{1}{2}\sigma_y^2\sop{}{y} \\
&+ \rho_{R \hatr} \sigma_R \sigma_{\hatr} \sqrt{R (1-R)\hatr}\cp{}{R}{\hatr}
+ \rho_{Rz}\sigma_R \sigma_z z\sqrt{R(1-R)} \cp{}{R}{z}
+ \rho_{\hatr z} \sigma_{\hatr} \sigma_z z \sqrt{\hatr} \cp{}{z}{\hatr} \nonumber \\
&+ \rho_{Ry}\sigma_R \sigma_y \sqrt{R(1-R)} \cp{}{R}{y}
+ \rho_{\hatr y} \sigma_{\hatr} \sigma_y \sqrt{\hatr} \cp{}{y}{\hatr}
+ \rho_{yz} \sigma_y \sigma_z z \cp{}{y}{z} \nonumber \\
&+ \kappa_R(\theta_R-R)\fp{}{R}
+ \kapr(\ther - \hatr) \fp{}{\hatr}
+ (r - \hatr) z \fp{}{z}
+ \kapy(\they - y) \fp{}{y}. \nonumber
\end{align}

The function $f(t, T, R,\hatr, y, z, 0) \equiv v(t, T, \bfX)$ solves another PDE
\begin{align} \label{PDE2}
\fp{v(t,T,\bfX)}{t} &+ {\cal L} v(t,T,\bfX) - r v(t,T,\bfX)
- \lambda \gamma_z z \fp{v(t,T,\bfX)}{z} \\
&+ \lambda \left[\hat{u}(t, T, \bfX) -
v(t, T, \bfX) \right] = 0. \nonumber
\end{align}
\noindent where based on \eqref{lambda}, $\lambda = e^y$.

The function $\hat{u}(t, T, \bfX)$ in \eqref{PDE2} is defined as follows.  Recall, that the function $u$ corresponds to states {\it after} default, and the function $v$ corresponds to states {\it before} default. At default the variables $z$ and $\hatr$ experience a jump proportional to the value of each variable. Therefore, we introduce a new vector of states $\bfX^+ := \{R, \hatr(1+\gamma_\hatr), y, z(1+\gamma_z)\}$. The above argumentation suggests that after the function $u(t,T,\bfX)$ is found by solving \eqref{PDE1}, this solution should be translated into the new point $\bfX^+$ by shifting  (jumping) the independent variables $\hatr$ and $z$. In other words, after this translation is done we obtain a new function $\bar{u}(t,T,\bfX^+) = u(t,T,\bfX)$. However, since in \eqref{PDE2} $v = v(t,T,\bfX)$, it is convenient to re-interpolate $\bar{u}(t,T,\bfX^+)$ back to $\bfX$, and we denote the result of this interpolation as $\hat{u}(t,T,\bfX)$. This is exactly the same method as was used in \cite{isvIJTAF}. Technically, this is similar to how option written on a stock paying discrete dividends are priced by using a finite-difference method, see e.g., \cite{fdm2000}.

\subsection{Boundary conditions} \label{bcSec}

The \eqref{PDE1} and \eqref{PDE2} form a system of backward PDEs which should be solved subject to some terminal conditions at $t=T$, and the boundary conditions set at the boundaries of the domain $(R, \hatr, y, z) \in [0,1] \times [0,\infty] \times [-\infty,0] \times[0,\infty]$. As the value of the bond price is usually not known at the boundary, a standard way is to assume that the second derivatives vanish towards the boundaries.

However, this is subject of the following consideration. As mentioned in \cite{ItkinCarrBarrierR3}, if the diffusion term vanishes at the boundary, the PDE degenerates to the hyperbolic one. Then to set the correct boundary condition at this boundary we need to check the speed of the diffusion term in a direction normal to the boundary versus the speed of the drift term. To illustrate, consider a PDE
\begin{equation} \label{oleinik}
C_t = a(x)C_{xx} + b(x)C_x + c(x) C,
\end{equation}
\noindent where $C= C (t,x)$ is some function of the time $t$ and the independent space variable $x \in [0,\infty)$, and $a(x), b(x)$, $c(x) \in \calC^{2}$ are some known functions  of $x$. Consider the left boundary to be at $x=0$. Then, as shown by \cite{OleinikRadkevich73}, no boundary condition is required at $x = 0$ if $\lim_{x \rightarrow 0}[b(x) - a_x(x)] \ge 0$. This means that the convection term at $x=0$ is flowing upwards and dominates as compared with  the diffusion term. An example of such consideration is the Feller condition as applied to the Heston model, \cite{Lucic2008}.

To clarify, no boundary condition at $x=0$ means that instead of the boundary condition at $x \rightarrow 0$ the PDE itself with coefficients $a(0), b(0), c(0)$ should be used at this boundary.

The first boundaries to check are $R \to 0$ and $R \to 1$ as the PDEs in \eqref{PDE1} and \eqref{PDE2} degenerate at those boundaries to the hyperbolic ones in the $R$ direction.
At $R \to 0$ we have
\begin{equation} \label{bcR0}
\lim_{R \rightarrow 0}\left[\kappa_R(\theta_R-R) - \frac{1}{2}\sigma_R^2 (1-2R)\right]  =
\kappa_R \theta_R - \frac{1}{2}\sigma_R^2.
\end{equation}
Thus, this is the Feller condition for the stochastic recovery $\calR$, i.e., no boundary condition is necessary at $R=0$ if $2 \kappa_R \theta_R /\sigma_r^2 \ge 1$. Similarly, at $R=1$ we have
\begin{equation} \label{bcR1}
\lim_{R \rightarrow 1}\left[\kappa_R(\theta_R-R) - \frac{1}{2}\sigma_R^2 (1-2R)\right]  =
\kappa_R (\theta_R-1) + \frac{1}{2}\sigma_R^2.
\end{equation}
Therefore, if $\kappa_R (\theta_R-1) + \frac{1}{2}\sigma_R^2 \le 0$, no boundary condition is required at $R=1$ \footnote{The sign of this inequality changes because the inflow flux at $R=1$ is oriented in the opposite direction to the inflow flux at $R=0$.}.

For $\hatr$ the condition similar to \eqref{bcR0} holds at $\hatr  = 0$. Other boundary conditions read
\begin{equation} \label{bc}
\sop{u}{\hatr}\Big|_{\hatr \uparrow \infty} =
\sop{u}{y}\Big|_{y \uparrow -\infty}
= \sop{u}{y}\Big|_{y \uparrow 0}
= \sop{u}{z}\Big|_{z \uparrow 0}
= \sop{u}{z}\Big|_{z \uparrow \infty} = 0.
\end{equation}

We assume that the default has not yet occurred  at the valuation time $t$, therefore, \eqref{bondPrice} reduces to
\begin{equation} \label{bondPrice1}
U_t(T, R, \hatr, y, z) = v(t, T, \bfX).
\end{equation}
Thus, it could be found by solving a system \eqref{PDE1}, \eqref{PDE2} as follows. Since the bond price defined in \eqref{payoff} is a sum of two terms, and our PDE is linear, it can be solved independently for each term. Then the solution is just a sum of the two.

\subsection{Solving the PDE for $w_t(T)$} \label{wtT}

The function $w_t(T)$ can be obtained by solving \eqref{PDE1}, \eqref{PDE2} \footnote{The PDEs remain unchanged since the model is same, and only the contingent claim $G(t,T,r,\hatr, y,z,d)$, which is a function of the same underlying processes, changes.} in two steps.

\paragraph {Step 1}
First, we solve the PDE in \eqref{PDE1} for $u(t, T, \bfX)$. Since this function  corresponds to $d=1$, it describes the evolution of the bond price {\it at or after} default, which implies the terminal condition at $t=T$
\begin{equation}
U(\bfX) = u(T, T, \bfX) = 0.
\end{equation}
This payoff does not assume any recovery paid at default, therefore, the bond expires worthless. By a simple analysis, one can conclude  that $u(t, T, \bfX) \equiv 0$ is the solution of our problem at $d=1$. Indeed, it solves the equation itself, and also obeys the terminal and boundary conditions. In other words, at this step we know the solution in closed form.

\paragraph {Step 2}  Based on the results of the first step,  we have $u(t, T, \bfX^+) \equiv 0$ in \eqref{PDE2}.

By the definition before \eqref{PDE2}, the function $v(t, T, \bfX)$ corresponds to the states with no default. Therefore, from \eqref{payoff} the payoff function $v(T,T,\bfX)$ reads
\begin{equation} \label{tc2}
v(T,T,\bfX) = z.
\end{equation}
This payoff is the terminal condition for \eqref{PDE2} at $t=T$. The boundary conditions, again are set as in Section~\ref{bcSec}.

Now the PDE in \eqref{PDE2} for $v(t,T,\bfX)$ takes the form
\begin{align}
\fp{v(t,T,\bfX)}{t} &+ {\cal L} v(t,T,\bfX) - (r + \lambda) v(t,T,\bfX)
- \lambda \gamma_z z \fp{v(t,T,\bfX)}{z} = 0,
\end{align}
\noindent which should be solved subject to the terminal condition in \eqref{tc2}.
Then, finally $w_t(T) = v(t,T,\bfX)$. It can be seen, that in case of no recovery, the defaultable bond price
does not depend on the jump in the foreign interest rate, but depends on the jump in the FX rate.

\subsection{Solving the PDE for $g_t(\nu)$} \label{gtT}

To compute the second part of the payoff in \eqref{payoff}, observe that the integral in \eqref{payoff} is deterministic in $\nu$. Therefore, it can be approximated by a Riemann--Stieltjes sum, i.e. the continuous interval $[t,T]$ is replaced by a discrete (e.g., uniform) grid with a small step $\Delta \nu = h$. Then
\begin{equation}\label{eq:Integral24}
\int_t^T g_t(\nu) d\nu \approx h \sum_{i=1}^N g_t(t_i),
\end{equation}
\noindent where $t_i = t + i h, \ i \in [0,N]$, $N = (T-t)/h$. It is important that each term of this sum can be computed independently by solving the corresponding pricing problem in \eqref{PDE1}, \eqref{PDE2} with the maturity $t_i, \ i \in [1,N]$. Obviously, solving many backward problems (one for each maturity) all together can be slow, but this is the consequence of the backward PDE approach. As mentioned in above, this can  be significantly improved by either using the forward approach, or by using parallelization. The later is fully possible since the solutions for every maturity $t_i, \ i \in [1,N]$ are independent.

Accordingly, for every maturity $t_i$ the function $g_t(T)$ again can be found in two steps.

\paragraph {Step 1}  At this step we solve \eqref{PDE1} using the boundary condition described in Section~\ref{bcSec} and the terminal condition
\begin{equation} \label{tcG}
u(T,T,X) = R z (1+\gamma_z)/T, \qquad T > 0,
\end{equation}
\noindent which is discussed in more detail in Appendix. At $T=0$ we set $g_T(T)\Big|_{T=0} = 0$.

\paragraph {Step 2} The function $v(t_i,T,\bfX)$ can be determined by solving \eqref{PDE2}.
Indeed, the values of parameters $\gamma_z, \gamma_\hatr$ are known, and the values of $\lambda$ (or $e^y$) are also set at all computational nodes (for instance, on a grid which is used to numerically solve the PDE problem in Step~1).

By definition before \eqref{PDE2}, the function $v(t_i,T,\bfX)$ corresponds to the states with no defaults. Accordingly, the recovery is not paid, which means that the terminal condition for this step vanishes
\begin{equation}
v(T,T,X) = V(X) = 0.
\end{equation}
Finally we set $g_t(T) = v(t,T,X)$. According to this structure, in case of non-zero recovery the price of a risky bond does depend on jumps in both FX and foreign IR rates.

\section{Backward PDE approach to price Quanto CDS} \label{bond2cds}

Here, similar to \cite{isvIJTAF} we use the model for the zero-coupon bond proposed in the previous sections and apply it to pricing of the CDS contracts. We remind that the CDS is a contract in which the protection buyer agrees to pay a periodic coupon to a protection seller in exchange for a potential cashflow in the event of default of the CDS reference name before the maturity of the contract $T$.

Below we follow the definitions and notation in \cite{isvIJTAF}. Namely, we assume that the CDS contract is settled at time $t$ and assures protection to the CDS buyer until time $T$. We consider CDS coupons to be paid periodically with the payment time interval $\Delta t$, and there will be totally $m$ payments over the life of the contract, i.e., $m \Delta t = T-t$. Assuming unit notional, this implies the following expression for the CDS coupon leg $L_c$, \cite{LiptonSavescu2014, BrigoMorini2005}
\begin{equation}
L_c = \mathbb{E}_t\left[\sum_{i=1}^{m} c B(t,t_i)\Delta t \m1_{\tau > t_i}\right],
\end{equation}
\noindent where $c$ is the CDS coupon, $t_i$ is the payment date of the $i$-th coupon, and $B(t,t_i) = B_t/B_{t_i}$ is the stochastic discount factor.

If the default occurs in between of the predefined coupon payment dates, there must be an accrued amount from the nearest past payment date till the time of the default event $\tau$. The expected discounted accrued amount $L_a$ reads
\begin{equation}
L_a = \mathbb{E}_t\left[c  B(t,\tau) (\tau - t_{\beta(\tau)}) \m1_{t < t_\beta(\tau) \le \tau < T}\right],
\end{equation}
\noindent where $t_{\beta(\tau)}$ is the payment date preceding the default event. In other words, $\beta(\tau)$ is a piecewise constant function of the form
\[ \beta(\tau) = i, \quad \forall \tau: \ t_i < \tau < t_{i+1}. \]
These cashflows are paid by the contract buyer and received by the contract issuer. The opposite expected protection cashflow $L_p$ is
\begin{equation}
L_p = \mathbb{E}_t\left[(1 - \calR_\tau)B(t,\tau)\m1_{t < \tau \le T}\right],
\end{equation}
\noindent where the recovery rate $\calR_\tau$  is unknown beforehand, and is determined at or right after the default, e.g., in court. To remind, in this paper in contrast to \cite{isvIJTAF} we assume the recovery rate to be stochastic.

We define the so-called \emph{premium} ${\cal L}_{pm} = L_c + L_a$ and \emph{protection} ${\cal L}_{pr} = L_p$ legs in the standard way, as well as define the CDS par spread $s$ as the coupon which equalizes these two legs and makes the CDS contract fair at time $t$. Under the domestic money market measure $\QM$ we need to convert the payoffs to the domestic currency and discount by the domestic money market account. Then $s$ solves the equation
\begin{align} \label{eq:CDSequation}
\sum_{i=1}^{m} & \mathbb{E}_t \left[s Z_T B(t,t_i)\Delta t \m1_{\tau > t_i}\right]
+ \mathbb{E}_t\left[s Z_T B(t,\tau)(\tau - t_{\beta(\tau)})\m1_{t<\tau<T}\right]
= \mathbb{E}_t\left[(1 - \calR_\tau)Z_\tau B(t,\tau)\m1_{t<\tau\leq T}\right].
\end{align}

In the spirit of \cite{ES2006} and \cite{BrigoSlide}, we develop a numerical procedure for finding the par spread $s$ from the bond prices. Consider each term in \eqref{eq:CDSequation} separately.

\paragraph{Coupons} For the coupon payment one has
\begin{align} \label{eqCoupon}
L_c &= \mathbb{E}_t \left[ \sum_{i=1}^{m} s Z_{t_i} B(t,t_i) \Delta t \m1_{\tau \ge t_i} \right] =  s\Delta t \sum_{i=1}^m \mathbb{E}_t \left[ Z_{t_i} B(t,t_i) \m1_{\tau \ge t_i} \right] = s \Delta t \sum_{i=1}^m w_t(t_i).
\end{align}
\noindent where $t_m = T$. Computation of $w_t(T)$ is described in Section~\ref{wtT}. In short, we first solve \eqref{PDE1} with the terminal condition $u(T,T,X) =0$, which can be solved analytically, and gives rise to $u(t,T,X) =0$. At the second step we solve numerically \eqref{PDE2} with the terminal condition $v(T,T,X) =z$.

Note, that as follows from the analysis of the previous section, $w_t(T)$ (and, respectively, the coupon payments)  does depend on the jump in the FX rate, but does not depend on the jumps in the foreign interest rate which is financially reasonable.

\paragraph{Protection leg} A similar approach is provided for the protection leg
\begin{align} \label{eqProtection}
L_p &= \mathbb{E}_t \left[(1-\calR_\tau) Z_{\tau} B(t,\tau)\m1_{t < \tau \le T} \right]
=   \int_{t}^{T} \mathbb{E}_t \left[(1-\calR_\nu)Z_\nu B(t,\nu) \m1_{\tau \in (\nu-d\nu,\nu]} \right] d\nu \\
&=  \int_{t}^{T} \bar{g}_t(\nu) d \nu, \qquad
\bar{g}_t(\nu) \equiv \mathbb{E}_t \left[(1-\calR_\nu)Z_\nu B(t,\nu) \m1_{\tau \in (\nu-d\nu,\nu]} \right]. \nonumber
\end{align}
Note, that computation of $\bar{g}_t(T)$ could be done in the same way as this is described in Section~\ref{gtT} for $g_t(T)$. This means that to find $\bar{g}_t(T)$ we first solve \eqref{PDE1} for $u(t,T,X)$ subject to the boundary conditions in Section~\ref{bcSec}, and the terminal condition
\begin{equation} \label{tcGbar}
u(T,T,X) = (1-R) z (1+\gamma_z)/T, \qquad T > 0,
\end{equation}
\noindent otherwise, we set $\bar{g}_T(T)\Big|_{T=0} = 0$. At the second step we solve \eqref{PDE2}, where $\hat{u}$ is computed by re-interpolating $u(t,T,X)$ found at the first step, subject to the boundary conditions in Section~\ref{bcSec}, and the terminal condition $v(T,T,X) = 0$. Then $\bar{g}_t(\nu) = v(t,\nu,X)$.

\paragraph{Accrued payments} For the accrued payment one has
\begin{align} \label{eqAccrued}
L_a &=  \mathbb{E}_t \left[ s Z_\tau B(t,\tau) (\tau - t_{\beta(\tau)}) \frac{\m1_{t < \tau < T}}{d\nu} \right]
= s \int_t^T \mathbb{E}_t \left[ Z_\nu B(t,\nu) (\nu - t_{\beta(\nu)}) \frac{\m1_{\tau \in (\nu-d\nu,\nu]}}{d\nu}  \right] d\nu \\
&= s \sum_{i=0}^{m-1} \Big\{\int_{t_i}^{t_{i+1}} (\nu - t_i) \mathbb{E}_t \left[ Z_\nu B(t,\nu) \frac{\m1_{\tau \in (\nu-d\nu,\nu]}}{d\nu}  \right] d\nu \Big\}
= s \sum_{i=0}^{m-1} \int_{t_i}^{t_{i+1}} (\nu - t_i)\tilde{g}_t(\nu) d\nu, \nonumber
\end{align}
\noindent where $t_0 \equiv t$, $t_m \equiv T$, and
\begin{equation} \label{tildeg}
\tilde{g}_t(\nu) \equiv \mathbb{E}_t \left[ Z_\nu B(t,\nu) \frac{\m1_{\tau \in (\nu-d\nu,\nu]}}{d\nu}  \right].
\end{equation}
Computation of $\tilde{g}_t(T)$ could be done in the same way as this is described in Section~\ref{gtT} for $g_t(T)$. This means that to find $\tilde{g}_t(T)$ we first solve \eqref{PDE1} for $u(t,T,X)$ subject to the boundary conditions in Section~\ref{bcSec}, and the terminal condition
\begin{equation} \label{tcGbartilde}
u(T,T,X) = z (1+\gamma_z)/T, \qquad T > 0,
\end{equation}
\noindent otherwise, we set $\tilde{g}_T(T)\Big|_{T=0} = 0$. At the second step we solve \eqref{PDE2}, where $\hat{u}$ is computed by re-interpolating $u(t,T,X)$ found at the first step, subject to the boundary conditions in Section~\ref{bcSec}, and the terminal condition $v(T,T,X) = 0$. Then $\tilde{g}_t(\nu) = v(t,\nu,X)$.

As was mentioned in Section~\ref{gtT}, both final integrals in \eqref{eqProtection}, \eqref{eqAccrued} are deterministic in $\nu$. Therefore, each one can be approximated by a Riemann--Stieltjes sum where the continuous interval $[t,T]$ is replaced by a discrete grid (e.g., uniform) with a small step $\Delta \nu = h$.

Having all necessary components in hands, we finally compute the CDS spread. Using new notation
\begin{align} \label{approx1}
A_i &= \int_{t_i}^{t_{i+1}} w_t(\nu) d\nu \approx h \sum_{k=1}^N w_t(\nu_k), \qquad
B_i = \int_{t_i}^{t_{i+1}} \bar{g}_t(\nu) d\nu \approx h \sum_{k=1}^N \bar{g}_t(\nu_k) \\
C_i &= \int_{t_i}^{t_{i+1}} \nu \tilde{g}_t(\nu) d\nu \approx h \sum_{k=1}^N \nu_k \tilde{g}_t(\nu_k), \qquad
D_i = \int_{t_i}^{t_{i+1}} t_i \tilde{g}_t(\nu) d\nu \approx h t_i \sum_{k=1}^N \tilde{g}_t(\nu_k) \nonumber \\
\nu_k &= t_i + k h, \quad k=1,\ldots,N, \quad h = (t_{i+1} - t_i)/N, \nonumber
\end{align}
\noindent then re-writing \eqref{eqCoupon}, \eqref{eqProtection} and \eqref{eqAccrued} in the form
\begin{align} \label{approx2}
L_p &= \sum_{i=1}^{m} B_i, \qquad
L_c = s  \sum_{i=1}^m A_i, \qquad
L_a = s \sum_{i=1}^{m} \left[ C_i - D_i \right],
\end{align}
and finally combining together \eqref{eq:CDSequation} and \eqref{approx2}, we obtain
\begin{align} \label{eqParSpread}
s &= \left\{\sum_{i=1}^{m} B_i \right\} \left\{\sum_{i=1}^{m} \left[ A_i + C_i - D_i\right]\right\}^{-1}.
\end{align}

\section{Solving backward PDEs using an RBF--FD method} \label{numMethod}

A key block in the described approach of computing a Quanto CDS spread is solving the backward PDEs in \eqref{PDE1}, \eqref{PDE2} subject to the corresponding terminal and boundary conditions. These PDEs are four-dimensional in space, and, therefore, some computational methods, e.g., finite-difference or finite element ones may suffer from the curse of dimensionality. On the other hand, a Monte-Carlo method is slow.

In this situation a rationale is to use localized meshless schemes such as, e.g., a radial basis function (RBF) generated finite-difference method (RBF--FD).  Within the last decade the localized RBF schemes have turned to be famous tools in solving various financial engineering problems (we refer a reader to \cite{Pettersson} for further discussion on technical issues of the RBF method, such as the number of discretization points and the structure of the discretization matrices optimal for solving these problems). That is why a flavor of the RBF method (a partition of unity method or RBF-PUM) was chosen in the recent work \cite{isvIJTAF} to solve a problem similar to that one in this paper.

As this is discussed in \cite{isvIJTAF}, the original formulation of the RBF methods is done based on global RBFs which leads to either ill-conditioned or dense matrices, and thus is computationally expensive, \cite{Fasshauer}. In \cite{isvIJTAF} this problem was eliminated by using the RBF-PUM. In more detail, see \cite{isvIJTAF} and references therein, and also a recent paper \cite{SM_RBF2017}.

In this paper we use another flavor of the localized RBF method known as RBF--FD, e.g. see \cite{FLYER201621, Bayona2017} and references therein. The RBF--FD is a combination of the localized RBF and the FD methods. Comparison of both methods is presented in \cite{SM_RBF2017} which illustrates high capability of those methods for solving PDEs to a sufficient accuracy within a reasonable time, while both methods exhibit similar orders of convergence, However, from the potential parallelization viewpoint the RBF--FD is, perhaps, more suitable.

The RBF--FD approximation is produced by applying the RBF interpolation at the finite local set of points. As the result, thus obtained matrices of derivatives are i) sparse, and ii) similar to that for the standard FD scheme, but have better convergence properties, \cite{Tolstykh}. Moreover,  the RBF--FD scheme is able to tackle irregular geometries and scattered node layouts. A particular flavor of the RBF--FD method used in this paper is described in \cite{Soleymani2018,FazItkin2019}. In this paper as the basis functions for the RBF method we use the Gaussian RBF $\phi(r_i)=e^{-\epsilon^2 r_i^2}$,  where $r_i=\|\mathbf{x}-\mathbf{x}_i\|_2$ denotes the Euclidean distance, and $\epsilon$ is the shape parameter. Other popular choices can also be used.

The discretization of the PDEs \eqref{PDE1}, \eqref{PDE2} is done first, along the spatial variables by using the RBF--FD methodology, and then  the method--of--lines (MOL) is used to finally reach a system of Ordinary Differential Equations (ODEs). The RBF--FD weights for approximation of the first and second derivatives in the above PDEs are obtained as in \cite{FazItkin2019}, and provide the second order of convergence.

\subsection{Time-integrator for \eqref{PDE1}, \eqref{PDE2}}

When solving the system of ODEs by using the MOL, for marching along the temporal variable we use a flavor of the time-stepping Runge-Kutta method. Namely, we employ the classical fourth order Runge-Kutta method  (RK4) meaning that the local truncation error is of the order of $\mathcal{O}(h^5)$, while the total accumulated error is of the order of $\mathcal{O}(h^4)$, where $h$ is the step of the method, \cite{Trott}.

Let us denote as $\mathbf{u}^\iota$ a numerical estimate for the exact solution $\mathbf{u}(t _\iota)$. Assume that we split the whole time interval $[0,T]$ into $k_\tau +1$ nodes with the step $\Delta_ t =T/k_\tau > 0$, so $ t _{\iota+1}= t _{\iota}+\Delta_  t $, $0\leq \iota\leq k_\tau$. Given the initial  condition $\mathbf{u}^0$, the RK4 scheme is formulated as follows, \cite{Butcher}:
\begin{align}\label{RK4}
\mathbf{u}^{\iota+1}&=\mathbf{u}^{\iota}+\frac{\Delta_  t }{6}\left(G_{1}+2G_{2}+2G_{3}+G_{4}\right), \nonumber \\
G_{1}&=H\left( t _{\iota},\mathbf{u}^{\iota}\right), \nonumber \\
G_{2}&=H\left( t _{\iota}+\frac{\Delta_  t }{2},\mathbf{u}^{\iota}+\frac{\Delta_  t }{2}G_{1}\right), \nonumber \\
G_{3}&=H\left( t _{\iota}+\frac{\Delta_  t }{2},\mathbf{u}^{\iota}+\frac{\Delta_  t }{2}G_{2}\right), \nonumber \\
G_{4}&=H\left( t _{\iota}+\Delta_  t ,\mathbf{u}^{\iota}+\Delta_  t G_{3}\right). \nonumber
\end{align}

From the implementation prospective, this method is a part of almost all standard packages including that of Wolfram Mathematica 12, which we use for doing all the calculations in this paper. It can be called by using the following snippet:
\centerline{}
\begin{verbatim}
Method -> {"FixedStep", "StepSize" -> temporalstepsize,
        Method -> {"ExplicitRungeKutta",
                    "DifferenceOrder" -> 4, "StiffnessTest" -> False}}
\end{verbatim}

Overall, based on the methodology described in Section~\ref{bond2cds}, computation of the CDS par spread in \eqref{eqParSpread} requires solving the PDE in \eqref{PDE1} two times, and the PDE in \eqref{PDE2} three times. An intermediate 4D interpolation of the solution $u(t,T,\textbf{X})$ to the solution $\hat{u}(t,T,\textbf{X}^+)$  is also required when proceeding from \eqref{PDE1} to \eqref{PDE2}, as this is explained in Section~\ref{zcbPrice}. The latter slows down the marching process when solving \eqref{PDE2}, if the number of discretization points is large.

\section{Experiments} \label{experiments}

Numerical experiments described in this Section are fulfilled in a way similar to that in \cite{isvIJTAF}. Also, to investigate quanto effects and their impact on the price of a CDS contract, we follow the strategy of \cite{isvIJTAF} and consider two similar CDS contracts. The first one is traded in the foreign economy, e.g., in Portugal, but is priced under the domestic risk-neutral $\QM$-measure, hence is denominated in our domestic currency (US dollars). To find the price of this contract our approach described in the previous sections is fully utilized.

The second CDS is the same contract which is traded in the domestic economy, and is also priced in the domestic currency. As such, its price can be obtained by solving the same problem as for the first CDS, but where the equations for the foreign interest rate $\hatr_t$ and the FX rate $Z_t$ are excluded from consideration. Accordingly, all related correlations which include index $z$ and $\hatr$ vanish, and the no-jumps framework is used.

However, the terminal conditions remain the same as in Section~\ref{bond2cds} as they are already expressed in the domestic currency\footnote{Alternatively, the whole four-dimensional framework could be used if one sets $z=1, \hatr = r, \gamma_z = \kapr = \sigr = \gamma_\hatr = 0$, and $\rho_{\cdot,z} = \rho_{\cdot,\hatr} = \rho_{z,\hatr} = 0$, where $\langle \cdot \rangle \in [R, z, y]$.}.

\subsection{Parameters of the model}

The default values of parameters used in our numerical experiments are same as in \cite{isvIJTAF} and are given below in Table~\ref{TabParam}. It is also assumed that for this default set all correlations are zero. If not stated otherwise, we use these values and assume the absence of jumps in the FX and foreign interest rates.
\begin{table}[H]
\begin{center}
\begin{tabular}{c c c c c c c c}
\hline
$R$ & $\kappa_R$ & $\theta_R$ & $\sigma_R$ & $\hat r$ & $\kapr$ & $\ther$ & $\sigr$ \\
0.45 & 0.0 & 0.1 & 0.0 & 0.03 & 0.08 & 0.1 & 0.08 \\
\hline
$y$ & $\kapy$ & $\they$ & $\sigy$ & $z$ & $\sigma_z$ & $T$ & $r$ \\
-4.089 & 0.0001 & -210 & 0.4 & 1.15 & 0.1 & 5 & 0.02\\
\hline
\end{tabular}
\caption{The default set of parameter values used in the experiments.}
\label{TabParam}
\end{center}
\end{table}
As it can be seen, by default we set $\kappa_R = \sigma_R = 0$, and $R=0.45$. This allows us to mimic a constant recovery rate that was used in \cite{isvIJTAF,Brigo}. Therefore, this default set provides almost the same conditions as the default set in \cite{isvIJTAF}. Here ``almost" means that the domestic interest rate $r$ in \cite{isvIJTAF} is stochastic, while here it is deterministic, but with the same initial value $r=0.02$. As follows from the results of \cite{isvIJTAF}, the domestic interest rate has a minor influence on the par spread, therefore, this comparison is eligible.

We also assume that coupons of the CDS contract are paid semi-monthly, thus the total number of coupon payments over the life of the CDS
contract is $m=120$.

In what follows, let us denote the CDS par spread found by using the first contract (traded in the foreign economy) as $s$, and the second one (traded in the domestic economy) as $s_d$. Hence, the quanto spread (or the quanto impact) is determined as the difference between these two spreads
\begin{equation}
\Delta s = s - s_d,
\end{equation}
\noindent which below, in agreement with the notation in \cite{Brigo}, we call the ``basis" spread.

\subsection{Parameters of the method}

The shape parameter $\epsilon$ of the Gaussian RBF is subject of a separate choice, see, e.g., \cite{Fornberg2}. In our numerical experiments it is always $\epsilon = 2h$.

As $(R, \hatr, y, z) \in [0,1] \times [0,\infty] \times [-\infty,0] \times[0,\infty]$, we truncate each semi-infinite or an infinite domain of definition sufficiently far away from the evaluation point, so an error brought by this truncation is relatively small. In particular, we use $0 \le \hatr \le 1$, $0 \le z \le 4$, $-6 \le y \le 0$. In case of jumps, the right boundary for $\hatr$ ($z$) is extended (truncated) by multiplying it by $1+\gamma_\hatr$ (or $1+\gamma_z$), respectively. Furthermore, the temporal step size for marching along time is chosen as $\Delta_t=0.05$.

We move the boundary conditions, defined in \eqref{bc}, to the boundaries of this truncated domain. Implementation-wise, the boundary conditions are explicitly incorporated into the pricing scheme. Hence, the latter can be implemented uniformly with no extra check that the boundary conditions are satisfied. As follows from Section~\ref{bcSec}, for the values of the model parameters given in Table~\ref{TabParam}, no boundary condition is required at the boundaries $R=0, R=1$ and $\hatr = 0$.

As long as $\gamma_z$ or $\gamma_\hatr$ are positive, the intermediate interpolation for constructing $\hat u$ (as this is described in Section~\ref{zcbPrice}) would encounter values outside of the truncated (computational) domain. In such a case instead of interpolation an extrapolation technique is applied by Mathematica 12.0 automatically. Accordingly, the command $\mathtt{Interpolation[]}$ is used throughout the code as much as required.

In all our numerical experiments we use [10, 10, 10, 10] discretization points uniformly spanned over the space $[R,\hatr,y,z]$. The temporal step size of the RK4 integrator is taken 0.0625. All computations were done at laptop having 16.0 GB of RAM, Windows 10, Intel(R) Core(TM) i7-9750H and SSD internal memory.

To speed up calculations as much as possible, we do the following. We set $\mathtt{PrecisionGoal -> 5}$, $\mathtt{AccuracyGoal -> 5}$  in our code to speed up the process of solving the large scale system of the discretized ODEs. Also all matrices are created as sparse arrays. The coefficient matrix (the right-hands matrix of the MOL method) is built by using the Kronecker products on tensor product grids. The boundary conditions are incorporated into this matrix in the same way. Finally, to compute the Riemann--Stieltjes sum in \eqref{eqParSpread} we use a built-in parallelization technique inside the Mathematica kernel to distribute the computational job over all available cores at our computer.
The typical elapsed time of computing the par spread using this procedure is roughly 210 seconds.

\subsection{Benchmarks}

Since in our test $\kappa_R = \sigma_R = 0$, and $R$ = const, computation of $s_d$ can be done by solving a one-dimensional problem where the only stochastic driver is $y$. Therefore, we also implemented our algorithm of solving \eqref{PDE1},\eqref{PDE2} by using a finite-difference method - the Crank-Nicholson scheme, \cite{fdm2000}. Thus found value of $s_d$ is compared with that obtained by using the full RBF-FD 4D algorithm where we set $z=1, \hatr = r, \gamma_z = \kapr = \sigr = \gamma_\hatr = 0$, and all the correlations vanish. The results are presented in Table~\ref{TabBench}, first two columns..
\begin{table}[!htb]
 \centering
 \begin{tabular}{|r|r|r|r|r|}
 \hline
 \multicolumn{2}{|c|}{$\kappa_y$ =0.0001, $\sigma_y$ = 0.4} & \multicolumn{3}{c|}{$\kappa_y = \sigma_y$ = 0} \\
 \hline
 4D & 1D & 4D & 1D & Asym \\
 \hline
  102.68    & 102.8   & 91.73   & 94.5   & 92.2 \\
 \hline
 \end{tabular}%
  \caption{Par spreads $s_d$ in bps obtained in our benchmark tests.}
 \label{TabBench}
\end{table}%

Also it is known that if the premium leg is paid continuously, and  the hazard rate $\lambda$ is constant, there exists a credit triangle relationship $s_d = \lambda (1-R)$. To mimic this in our calculations, we set $\kappa_y = \sigma_y = 0$ and repeat the calculations of $s_d$ by using the 4D RBF-FD and 1D FD algorithms. These results are compared with the credit triangle value and also presented in Table~\ref{TabBench}, the right two columns. Also, the obtained values are close to that reported in \cite{Brigo}. It can be seen, that the 4D implementation produces the results which are close to the benchmarks values (they differ by no more than 0.5\%).

In the absence of jumps, i.e., when $\gamma_z  = \gamma_\hatr = 0$, the computed value of $s$  corresponding to the model parameters in Table~\ref{TabParam} is 93.22 bps. Thus, it deviates from $s_d$ by a basis of -9.46 bps. In other words, with no jumps or correlations the quanto effect on the CDS spread is small (this was also observed in \cite{isvIJTAF,Brigo}).

\begin{figure}[h!]
\captionsetup[subfigure]{justification=centering}	
\centering
\subcaptionbox{The behavior of $\Delta s$ as a function of $\gamma_\hatr$.\label{ds_gamma:a}}
{\includegraphics[height=2.in]{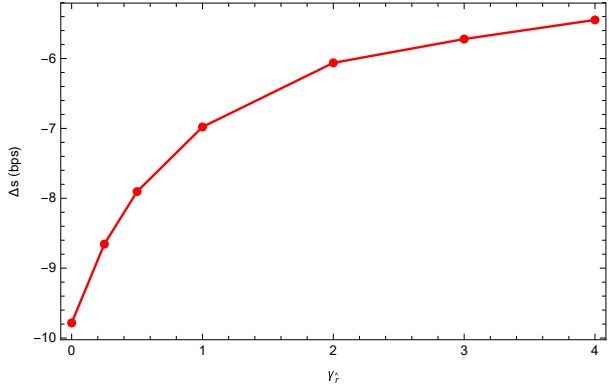}}
\hfill
\subcaptionbox{The behavior of $\Delta s$ as a function of $\gamma_z$.\label{ds_gamma:b}}
{\includegraphics[height=2.in]{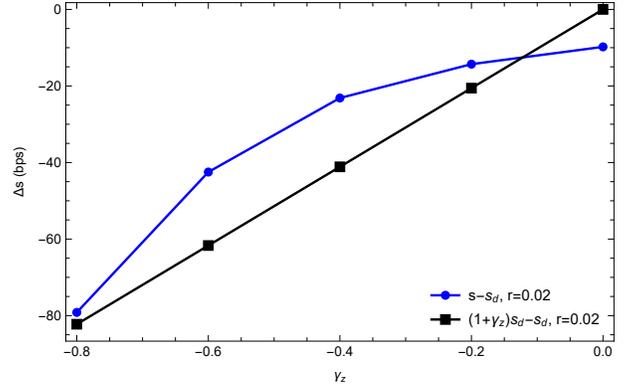}}
\caption{The influence of the jump-at-default amplitude $\gamma_\hatr$ and $\gamma_z$ on the five-year CDS basis spread.}
\label{ds_gamma}
\end{figure}

In Fig.~\ref{ds_gamma} the influence of the jump-at-default amplitude $\gamma_z$ (Fig.~\ref{ds_gamma:b}) and $\gamma_\hatr$ (Fig.~\ref{ds_gamma:a}) on the basis CDS spread is presented. These plots are similar to those in Fig.~4 in \cite{isvIJTAF}.
In our case the influence of $\gamma_\hatr$ is a bit less significant (by about 5 bps), and the influence of $\gamma_z$ is also less significant (-80 bps here vs -300 bps at $\gamma_z = -0.8$). This can be explained by the difference in $s_d$ (102.68 bps here vs 365 bps in \cite{isvIJTAF} which, in turn, can be attributed to the stochastic interest rate $r$ in \cite{isvIJTAF}  vs a constant interest rate in this paper).

The financial meaning of the behavior of $\Delta s$ with changes in $\gamma_\hatr$ and $\gamma_z$ is explained in \cite{isvIJTAF}. In the case when the credit triangle approximation is accurate, it is obtained in \cite{Brigo} that $s \approx (1+\gamma_z)s_d$.  In words, this means that the CDS spread expressed in the foreign currency is approximately proportional to the reference USD spread, and the coefficient of  proportionality is $(1+\gamma_z)$. In turn, the coupon payments expressed in the foreign currency are lower. In Fig.~\ref{ds_gamma:b} we also present this dependence to see that the results obtained by using our model align with the above approximation at $\gamma_z \approx -0.8$ and  $\gamma_z \approx -0.1$, but slightly deviate from this straight line at the intermediate values of $\gamma_z$.

Overall, these benchmark tests justify that our numerical approach provides reasonable values and behavior of the basis spread at some typical values of the model parameters. Therefore, further on we use the proposed model and numerical method to investigate the impact of stochasticity of the recovery rate on the quanto effect.

\subsection{Results}

To recall, in all our calculations presented in this Section we use the default set of parameters given in Table~\ref{TabParam} unless different values of the parameters are explicitly specified. This also assumes no jumps-at-default ($\gamma_z = \gamma_\hatr = 0$) in the default set of parameters.
\begin{figure}[h!]
\centering
\includegraphics[width=0.6\linewidth]{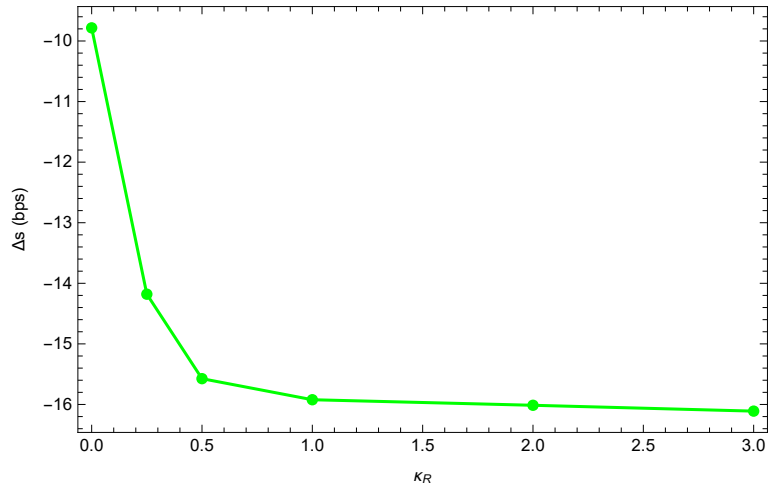}
\caption{The influence of the mean-reversion rate $\kappa_R$ on the five-year CDS basis spread.}
\label{ds_kappaR}
\end{figure}

In the first test we look at the influence of $\kappa_R$ on the basis spread. These results are presented in Fig.~\ref{ds_kappaR}. Note, that changes in $\kappa_R$ also affect the value of $s_d$. It can be seen that the increase in $\kappa_R$ results in the negative increase of $\Delta s$. In other words, while the decrease in $\gamma_z$ (the jump in the FX rate) gives rise to the decrease of $\Delta s$, the increase of $\kappa_R$ has the same effect. This is not obvious in advance, since the increase of $\kappa$ also increases the recovery rate, and hence the CDS spread value decreases. However, it also decreases $s_d$. Finally, it turns out that $s$ drops down faster than $s_d$, and, therefore,  the difference $\Delta s = s-s_d$ decreases.

We would also expect a pronounced influence of the recovery volatility $\sigma_R$ on the value of the basis spread. However, surprisingly we didn't find this influence. In other words, in the absence of correlations, the magnitude of the second derivatives $\sop{u}{R}$ and
$\sop{v}{R}$ is very small.

However, this volatility is also a part of the various mixed derivative terms. Therefore, our next test is to look at the influence of $\sigma_R$ on $\Delta s$ when the corresponding correlations do not vanish. By financial meaning, at default the FX rate drops down  by factor $1+\gamma_z$. Therefore, the decrease in $z$ signals about a lost of creditworthiness As such, the recovery rate should also drop down, hence the correlation $\rho_{z,R}$ should be positive. When the default log-intensity $y$ increases, the default becomes more probable, and the recovery should decrease. Hence, $\rho_{y,R}$ should be negative. Similarly, as we assume that at default the foreign interest rate increases, $\rho_{\hatr,R}$  should be negative.
\begin{figure}[h!]
\captionsetup[subfigure]{justification=centering}	
\centering
\subcaptionbox{The behavior of $\Delta s$ as a function of $\sigma_R$ for $\rho_{z,R} > 0$.\label{ds_sigmaR:a}}
{\includegraphics[height=2in]{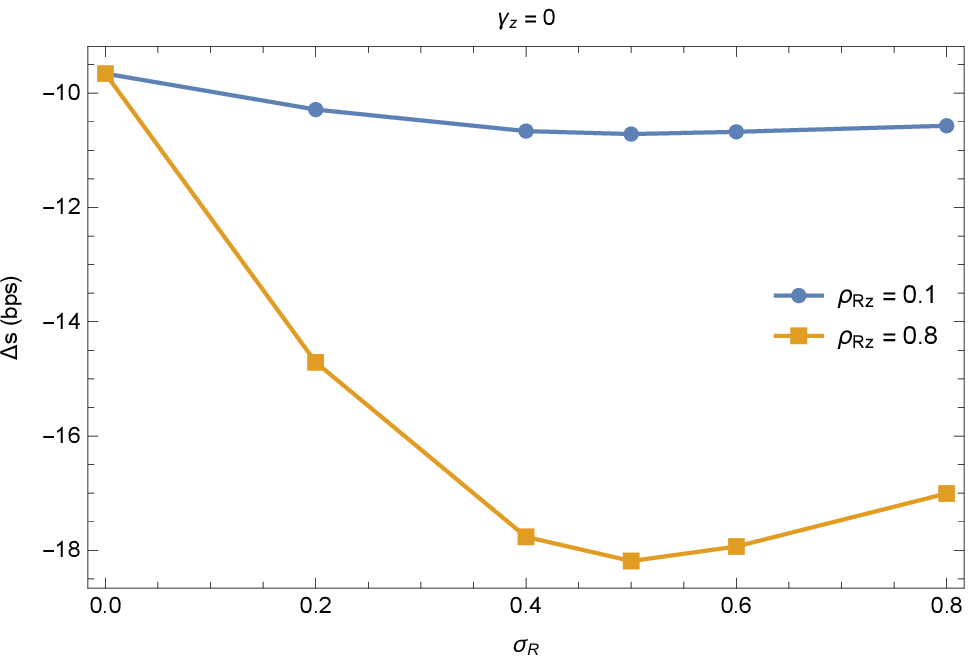}}
\hfill
\subcaptionbox{The behavior of $\Delta s$ as a function of $\sigma_R$ for $\rho_{y,R} < 0$.\label{ds_sigmaR:b}}
{\includegraphics[height=2in]{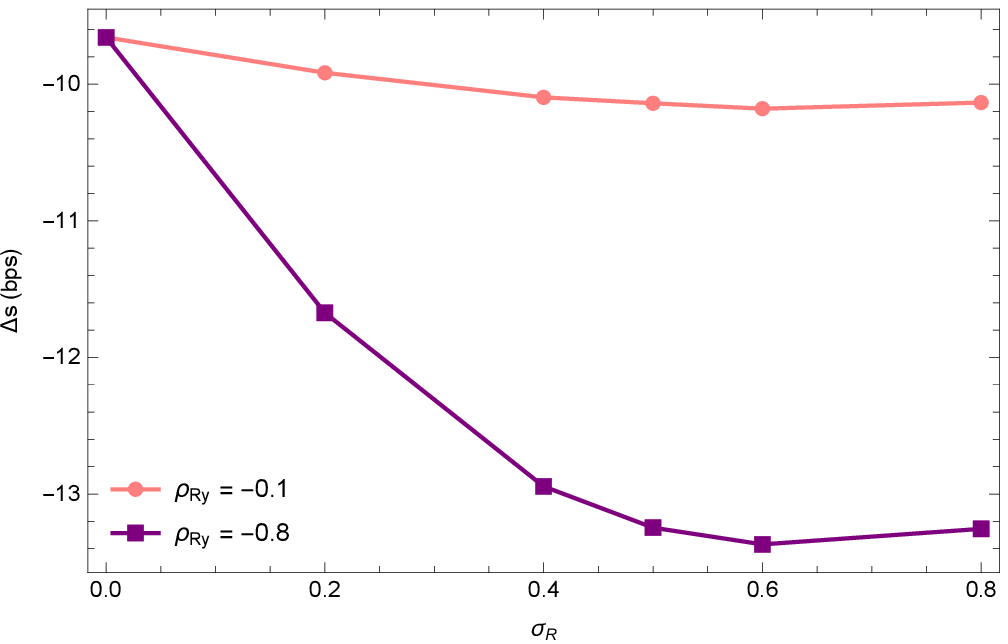}}
\caption{The influence of $\sigma_R$ on the five-year CDS basis spread when
$\rho_{z,R}$ and $\rho_{z,y}$ do not vanish.}
\label{ds_rhoR_1}
\end{figure}

In Fig.~\ref{ds_rhoR_1} we present the influence of $\sigma_R$ on the five-year CDS basis spread when the correlation
$\rho_{z,R}$ takes the values 0.1 and 0.8, and $\rho_{z,y}$ takes the values -0.1 and -0.8. It can be seen that with high negative correlation $\rho_{y,R}$, the increase in $\sigma_R$ results in a slow negative increase of $\Delta s$ (Fig.~\ref{ds_sigmaR:b}), and the influence of $\sigma_R$ at non-zero positive correlation $\rho_{z,R}$ looks to be similar and a bit more pronounced.
\begin{figure}[h!]
\centering
\includegraphics[width=0.6\linewidth]{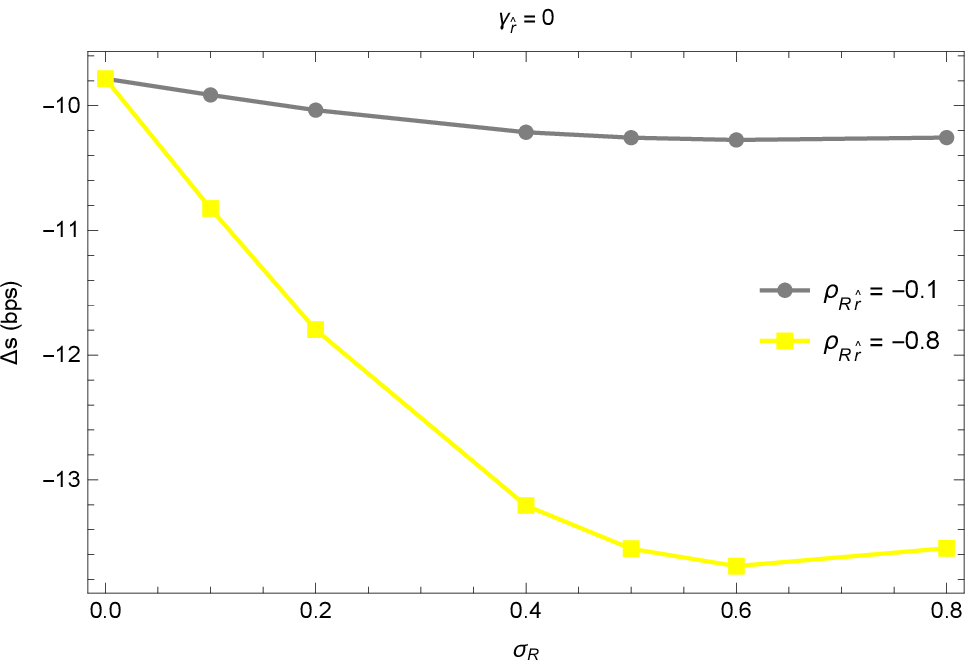}
\caption{The influence of $\sigma_R$ on the five-year CDS basis spread when $\rho_{R,\hatr}$ takes the values -0.1 and -0.8.}
\label{ds_rhoR_2}
\end{figure}

Same is true when the correlation $\rho_{R,\hatr}$ takes the values -0.1, -0.8 as this is presented in Fig.~\ref{ds_rhoR_2}. Thus, among all correlations the biggest impact can be attributed to $\rho_{z,R}$.

\begin{figure}[h!]
\centering
\includegraphics[width=0.6\linewidth]{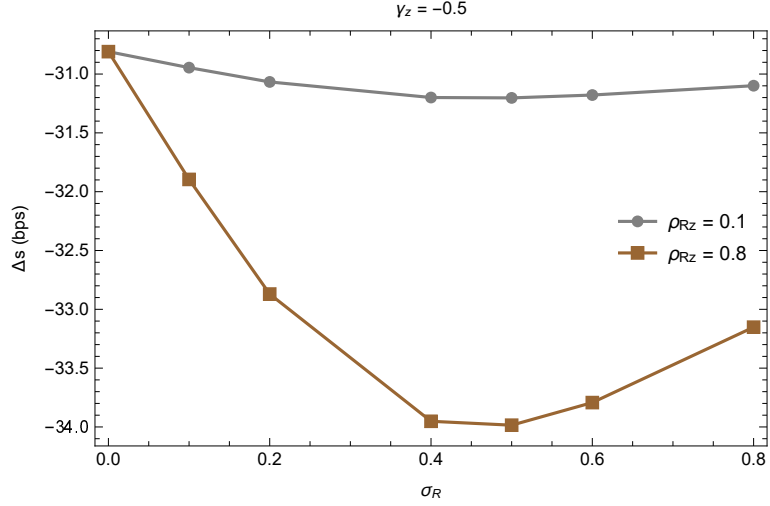}
\caption{The influence of $\sigma_R$ on the five-year CDS basis spread when $\rho_{R,z}$ takes the values -0.1 and -0.8, and $\gamma_z = -0.5$.}
\label{ds_rhoRz_gamma}
\end{figure}

The above results have been obtained with no jumps. Obviously, combining jumps and non-zero correlations will produce a joint effect.
This is illustrated in Fig.~\ref{ds_rhoRz_gamma} where we present the results of the same test as in Fig.~\ref{ds_sigmaR:a} but with $\gamma_z = -0.5$. Despite the value of the basis changes (due to the change in $\gamma_z$) the influence of $\sigma_R$ remains to be small.

\begin{figure}[h!]
\centering
\includegraphics[width=0.6\linewidth]{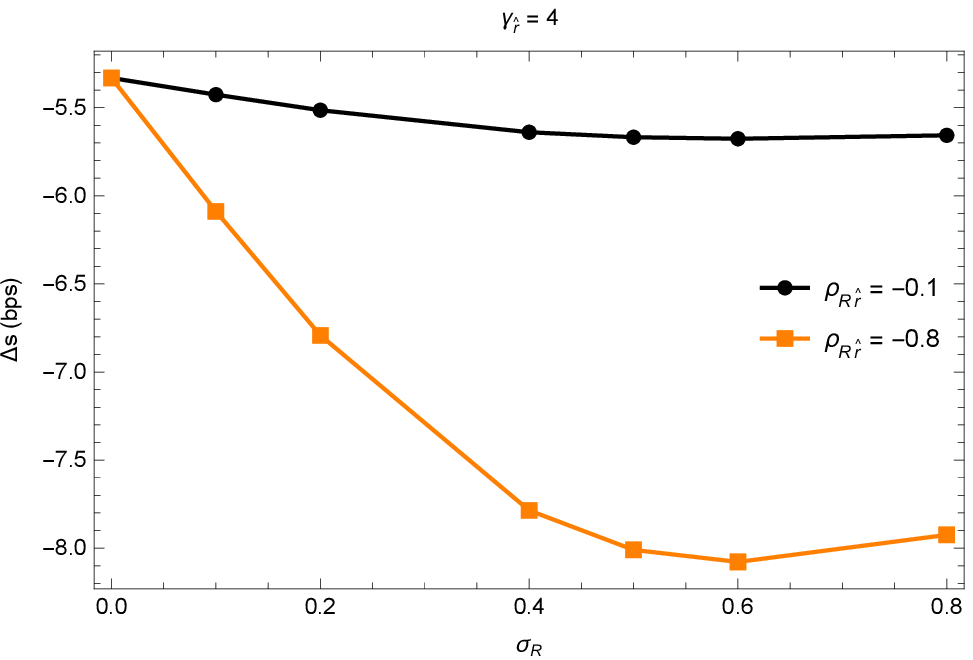}
\caption{The influence of $\sigma_R$ on the five-year CDS basis spread when $\rho_{R,\hatr}$ takes the values -0.1 and -0.8, and $\gamma_\hatr = 4$.}
\label{ds_rhoRr_gamma}
\end{figure}

It turns out, that the same is true when one adds jumps to the foreign interest rate. This has a relatively minor effect on the dependence of $\Delta s $ on $\sigma_R$, just the curves move up by about 5 bps due to the change in $\gamma_\hatr$. This behavior is presented in Fig.~\ref{ds_rhoRr_gamma}.

\section{Conclusion}\label{sec:Conclusion}

In this model we introduce another modification of the model for pricing Quanto CDS that was developed, first in \cite{Brigo}, and then in \cite{isvIJTAF}. Similar to \cite{isvIJTAF}, the presented model operates with four stochastic factors, which are the hazard rate (more rigorously, the log-intensity of default), the foreign exchange and interest rates and the recovery rate. The latter is new as in both \cite{Brigo, isvIJTAF} it was assumed to be constant. By using a PDE approach we derive the corresponding systems of PDEs for pricing both the risky bond and the CDS spread, also similar to how this is done in \cite{BieleckiPDE2005,isvIJTAF}. We make a careful analysis of the suitable boundary and initial conditions, and show at which boundaries the boundary conditions should be omitted (similar to the well-known Feller condition).

To solve these systems of 4D PDEs we use a different flavor of the RBF method which is a combination of localized RBF and finite-difference methods, and is known in the literature as RBF--FD. As compared with the corresponding Monte Carlo or FD methods, in our four-dimensional case this method provides high accuracy and uses much less resources. We use the backward PDE approach which is slower than the corresponding forward approach, but is suitable for parallelization.

Before investigating the influence of the stochastic recovery rate on the Quanto CDS spread, we benchmark our model and method against the results of \cite{Brigo, isvIJTAF}. We show that our approach described in this paper provides close results to those in \cite{Brigo,isvIJTAF} when the same set of parameters (the constant recovery, values of parameters, etc.) is used. This allows the authors to inherit and incorporate results of \cite{isvIJTAF} into our framework, and be concentrated just of the new features of the model.

As shown in \cite{Brigo,isvIJTAF}, these flavors of models are capable to qualitatively explain the differences in the marked values of CDS spreads traded in foreign and domestic economies (accordingly, they are denominated in the foreign and domestic (USD) currencies). To recall,
both CDS contracts traded in the different economies can also be priced in the same currency, e.g., in USD. In this case the market demonstrates a spread between these two prices which is called the quanto spread. The existence of the quanto spread, to a great extent, can be explained by the devaluation of the foreign currency. Financially, this means a much lower protection payout if converted to the US dollars.

As compared with \cite{Brigo,isvIJTAF}, in this paper we analyze the impact of the stochastic recovery rate on the basis spread $\Delta s$. We found that changes in the recovery mean-reversion rate $\kappa_R$ moderately affect the value of $\Delta s$ even with no jump-at-default in the FX or foreign interest rate, while the changes in the recovery volatility $\sigma_R$ almost have no impact. The influence becomes more perceptible  if one takes into account various correlations of the recovery rate, namely $\rho_{z,R}, \rho_{y,R}, \rho_{\hatr,R}$.  Also, for non-zero correlations the dependence of $\Delta s$ on $\sigma_R$ becomes observable. Overall, in our setting  the maximum impact is about 10 bps, or 10\%.

These results have been obtained with no jumps. Obviously, it is expected that combining jumps and non-zero correlations will produce a joint effect. However, we found that while the jumps in $z$ definitely move the curves down (so the basis spread negatively increases) the dependences of $\Delta s$ on $\sigma_R$ and $\kappa_R$ remain almost same as at $\gamma_z = 0$. A similar picture can be observed for jumps in $\hatr$. Thus, we observe the influence of the stochastic recovery on the quanto spread not exceeding 10\%. This, however, is almost the same order of influence as the jumps at default in $\hatr$ produce, but they work in the different directions.

The proposed model together with the obtained results are new and represents the main contribution of this paper. Also, despite the numerical method has been already described in the previous papers of the authors (both joint and separate), its application to solving a system of the 4D PDEs derived in this paper is new (also new boundary conditions are used as compared with, e.g., \cite{FazItkin2019}). Finally, in this paper the method was parallelized to achieve the best efficiency.

At the end we have to mention that calibration of this model is computationally expensive because of a large number of the model parameters. As explained in \cite{isvIJTAF}, it can be split into few steps, where at every step we calibrate just a subset of parameters, and each set could be calibrated by using different financial instruments.

\section*{Acknowledgments}
We thank Damiano Brigo and Victor Shcherbakov for valuable discussions on the impact of the recovery rate and the RBF method. We assume full responsibility for any remaining errors.

\vspace{0.5in}
\appendix

\section{The terminal condition for $g_T(T)$}

Assume that $T > 0$. Otherwise, we set $g_T(T)\Big|_{T=0} = 0$. Using the definition of $g_t(T)$ in \eqref{payoff}, we can set $t=T$, and by conditioning on $R_t = R, \hatR_t = \hatr, Z_t = z, Y_t = y, d=1$ obtain
\begin{align} \label{termG}
g_t(\nu)\Big|_{t=T, \nu=T} &= \mathbb{E}_T \left[ \calR_T B(T-dT,\tau) Z_T \frac{\m1_{\tau \in (T-dT,T]}}{dT} \right] = \frac{R z }{dT} \mathbb{E}_T \left[\m1_{\tau \in (T-dT,T]} \right] \\
&= R z \lambda(T) = R z e^y, \nonumber
\end{align}
\noindent see \cite{Schonbucher2003}, Section~3.2.

On the other hand, consider \eqref{PDE1}. Its formal solution using the operator exponential can be represented in the form
\begin{equation} \label{solU}
u(t,T,\bfX) = e^{\int_t^T ({\cal L} - r) d\tau} u(T,T,\bfX).
\end{equation}
Now, assume that $\gamma_z = \gamma_\hatr = 0$ (so $\hat u \equiv u$), and re-write \eqref{PDE2} in the form
\begin{align} \label{PDEv}
\fp{v(t,T,\bfX)}{t} &+ {\cal L}_1 v(t,T,\bfX) = - \lambda u(t, T, \bfX), \\
{\cal L}_1 &\equiv {\cal L} - (r+\lambda). \nonumber
\end{align}
With allowance for \eqref{solU}, and the terminal condition $v(T,T,\bfX) = 0$,  the formal solution of this equation reads
\begin{align} \label{solV}
v(t,T,\bfX) = e^{\int_t^T ({\cal L} - r - \lambda) d\tau}\left(e^{ \lambda t}-1\right) u(T,T,\bfX).
\end{align}

Since $g_t(T) = v(t,T,\bfX)$, from \eqref{solV} we have
\begin{equation} \label{finU}
 g_T(T) = \left(e^{- \lambda T}-1\right) u(T,T,\bfX).
 \end{equation}
 Comparing \eqref{termG} and \eqref{finU}, one can see that we need to choose
\[ u(T,T,\bfX) = R z \lambda\left(e^{ \lambda T}-1\right)^{-1}. \]
However, since in our experiments $\lambda T < 0.1$ we can rewrite this condition in the form
\begin{equation} \label{finU1}
u(T,T,\bfX) \approx \frac{R z \lambda}{\lambda T} = \frac{R z}{T}.
\end{equation}

Finally, we need to take into account that the dynamics of $Z_t$ in \eqref{dzJump} implies that when the default occurs, the value of $Z_{\tau^-}$ jumps proportionally to itself, i.e., $Z_\tau = Z_{\tau^-}(1+\gamma_z)$. Thus, we get the additional multiplier $1+\gamma_z$ in the RHS of \eqref{finU1}. Thus obtained terminal condition is now used in \eqref{tcG}.

\vspace{0.5in}
\section*{References}

\end{document}